\documentclass[]{interact}
\usepackage[utf8]{inputenc}
\usepackage[T1]{fontenc}
\usepackage[normalem]{ulem}
\usepackage{hyperref}
\usepackage{url}

\usepackage{overpic}

\usepackage{xcolor}

\newcommand{\ch}[1]{{#1}TiO$_3$}

\usepackage{epstopdf}
\usepackage[caption=false]{subfig}

\usepackage[numbers,sort&compress]{natbib}
\bibpunct[, ]{[}{]}{,}{n}{,}{,}

\theoremstyle{plain}

\theoremstyle{definition}

\theoremstyle{remark}

\begin{document}


\title{BaTiO$_3$ -- SrTiO$_3$ composites: a microscopic study on paraelectric cubic inclusions}

\author{
\name{Sheng-Han Teng\textsuperscript{a}\thanks{CONTACT Sheng-Han Teng. Email: sheng-han.teng@ruhr-uni-bochum.de}, Chinwendu Anabaraonye\textsuperscript{a,b,c} and Anna Gr\"unebohm\textsuperscript{a}}
\affil{\textsuperscript{a}Interdisciplinary Centre for Advanced Materials Simulation (ICAMS) and Center for Interface-Dominated High Performance Materials (ZGH), Ruhr-University Bochum, Universit\"atsstr 150, 44801 Bochum, Germany\\
\textsuperscript{b}Institute for Physical Chemistry, University of Münster, Corrensstraße 28/30, 48149 Münster, Germany\\
\textsuperscript{c}International Graduate School for Battery Chemistry, Characterization, Analysis, Recycling and Application (BACCARA), University of Münster, Corrensstraße 40, 48149 Münster, Germany}
}

\maketitle

\begin{abstract}
Composites of ferroelectric and paraelectric perovskites have attracted a lot of attention due to their application potential in energy storage as well as novel computing and memory devices.
So far the main focus of research has been on superlattices and ferroelectric particles in a paraelectric matrix, while the impact of paraelectric inclusions on a ferroelectric matrix is surprisingly underrepresented. To close this gap in knowledge we perform molecular dynamics simulations using an $ab\ initio$ derived effective Hamiltonian for BaTiO$_3$--SrTiO$_3$ and reveal the dependency of phase stability and phase transitions on the size and distances of paraelectric inclusions. 
We discuss how the combination of compressive strain and depolarization fields at the SrTiO$_3$ interfaces induces large local polarization, complex domain structures and coexisting phases as well as diffuse phase transitions and reduced coercive fields.
\end{abstract}

\begin{keywords}
Ferroelectric composites; BaTiO$_3$; SrTiO$_3$ inclusions; Phase transition; Phase coexistence; Domain structures;
\end{keywords}


\section{Introduction}
\label{sec:Introduction}
Ferroelectric perovskite oxides are widely utilized in modern applications ranging from capacitors, sensors, actuators as well as energy harvesting and storage devices to new computing concepts and memories \cite{acostaBaTiO3basedPiezoelectricsFundamentals2017, grunebohmInterplayDomainStructure2021, natafUsingOxidesCompute2024}.
Recently, composite materials with ferroelectric and paraelectric volume fractions arose attention due to the emergence of nanodomains and polar topologies such as vortices and skyrmions \cite{lisenkovUnusualStaticDynamical2009, yadavObservationPolarVortices2016,  grunebohmInterplayDomainStructure2021} as well as negative capacitance and antiferroelectric-like behavior
\cite{walterStrainControlLayerresolved2020, aramberriFerroelectricParaelectricSuperlattices2022, liuPhaseFieldSimulationsTunable2022}.

So far, the main focus on composites has been on 2D superlattices. 
In this layered geometry, misfit strain can change the ferroelectric transition temperatures, enhance the spontaneous polarization by an order of magnitude 
\cite{grunebohmInitioPhaseDiagram2015},  and induce ferroelectric phase transitions in otherwise paraelectric layers \cite{neatonTheoryPolarizationEnhancement2003, wangLargePolarization502024}. 
Additionally, it has been shown that thicknesses of the layers and the material choice can affect phase stabilities as well as chirality, stability and dynamics of polar structures \cite{jiangStabilityChiralPolarization2022, daiStrainEffectsStability2023}. These findings have been related to the interplay of strain and depolarization at the internal material interfaces.
Due to their simple geometry, superlattices are now routinely fabricated \cite{wangLargePolarization502024} and are accessible by \textit{ab\ initio} and molecular dynamics (MD) \cite{neatonTheoryPolarizationEnhancement2003, dimouPinningDomainWalls2022,walterStrainControlLayerresolved2020} and phase-field simulations \cite{jiangStabilityChiralPolarization2022, liuPhaseFieldSimulationsTunable2022, daiStrainEffectsStability2023}.

Composites with other morphologies are so far underrepresented in literature and are computationally demanding in microscopic simulations.
Coarse-grained MD simulations based on the effective Hamiltonian by Zhong et al.\ \cite{zhongPhaseTransitionsBaTiO31994} and phase-field simulations have been successfully used to address nanoparticles and nanowires. Thereby the main focus has been on ferroelectric inclusions in a paraelectric matrix. 
It has been shown that these nanowires can host a skyrmionic state that depends on the radius of the nanowires \cite{nahasDiscoveryStableSkyrmionic2015} and that may improve the energy storage capacity of the material \cite{jiangEnergyStorageProperties2023a}. Furthermore, dipole vortices in nanodots have been demonstrated and their dependency on volume fraction and the interaction with the
matrix has been reported \cite{prosandeevPropertiesFerroelectricNanodots2006, lichLowfieldEnergyStorage2024}.
 \ch{Ba}--\ch{(Ba,Sr)} composites have successfully been synthesized by spark plasma sintering \cite{takeuchiPreparationBaTiO3SrTiO32003}.
To the best of our knowledge, paraelectric nanoparticles embedded in a ferroelectric matrix are however surprisingly unexplored in simulations. 

In this work, we want to close this gap in knowledge and study cubic inclusions of \ch{Sr} embedded in a \ch{Ba} matrix.
On the one hand, \ch{Ba} is a prototypical and environmentally friendly perovskite with three first-order phase transitions which is sensitive to strain and chemical pressure \cite{acostaBaTiO3basedPiezoelectricsFundamentals2017,grunebohmInitioPhaseDiagram2015,wangChemicalPressureModulatedBaTiO3Thin2021}. 
On the other hand, \ch{Sr} is a quantum paraelectric perovskite with a tetragonal phase and antiferrodistortive (AFD) rotations below 105~K \cite{zhongCompetingStructuralInstabilities1995}. 
We perform simulations with an $ab\ initio$ derived effective Hamiltonian to screen different densities and sizes of these inclusions to narrow down the parameter space of interest for future experimental studies.
We show that while \ch{Sr} inclusions smaller than 3.2~nm have little effect on the ferroelectric phases of \ch{Ba} but may already reduce the coercive field, dense inclusions larger than 7.2~nm  induce diffuse phase transitions and stabilize heterogeneous multi-domain structures with coexisting phases in a broad temperature range already for an overall \ch{Sr} concentration of less than 5\%.


\section{Methods}
\label{sec:Methods}
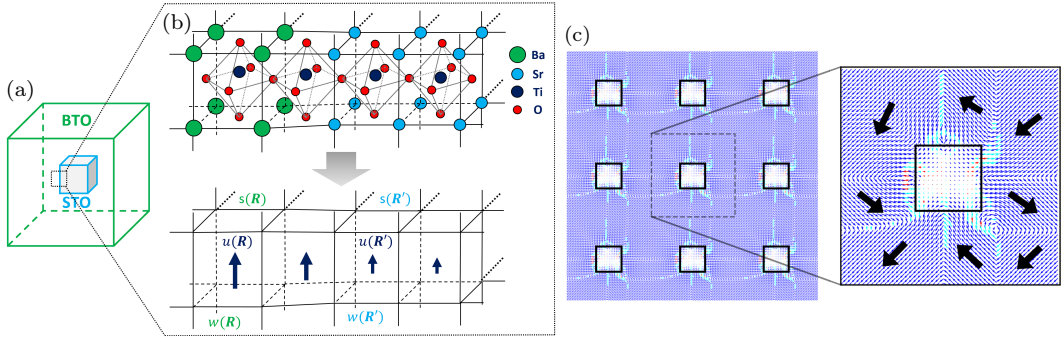
\begin{figure}[!htb]
    \centering
    \resizebox{.48\linewidth}{!}{
    \begin{Overpic}[abs]{\begin{tabular}{p{.43\textwidth}}\vspace{.18\textheight}\\\end{tabular}}
        \put(0,0){\includegraphics[width=0.5\textwidth]{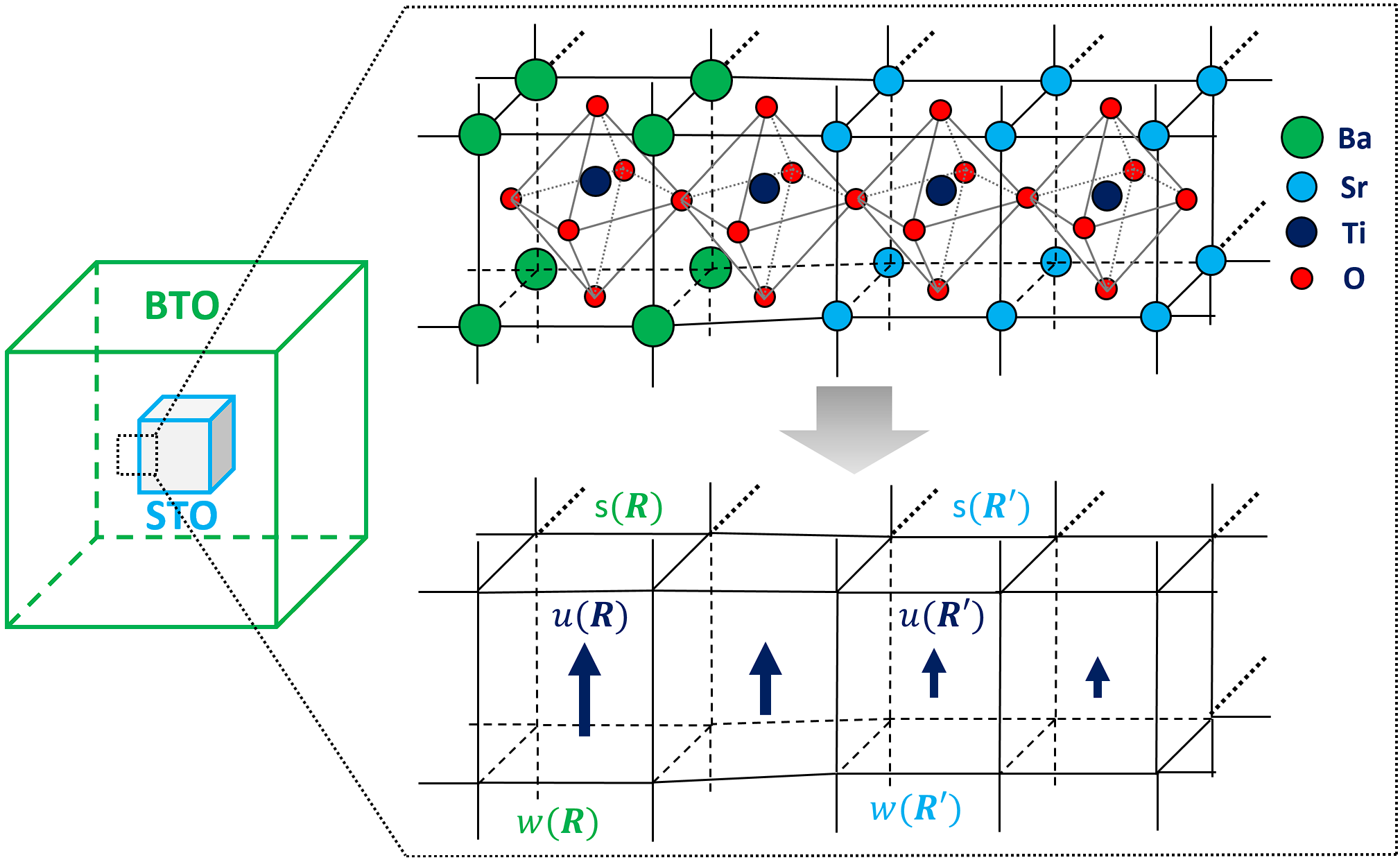}}
        \put(0,90){\scriptsize(a)}
        \put(58,115){\scriptsize(b)}
    \end{Overpic}
    }
    \resizebox{.48\linewidth}{!}{
    \begin{Overpic}[abs]{\begin{tabular}{p{.45\textwidth}}\vspace{.18\textheight}\\\end{tabular}}
        \put(2,5){\includegraphics[width=.5\linewidth]{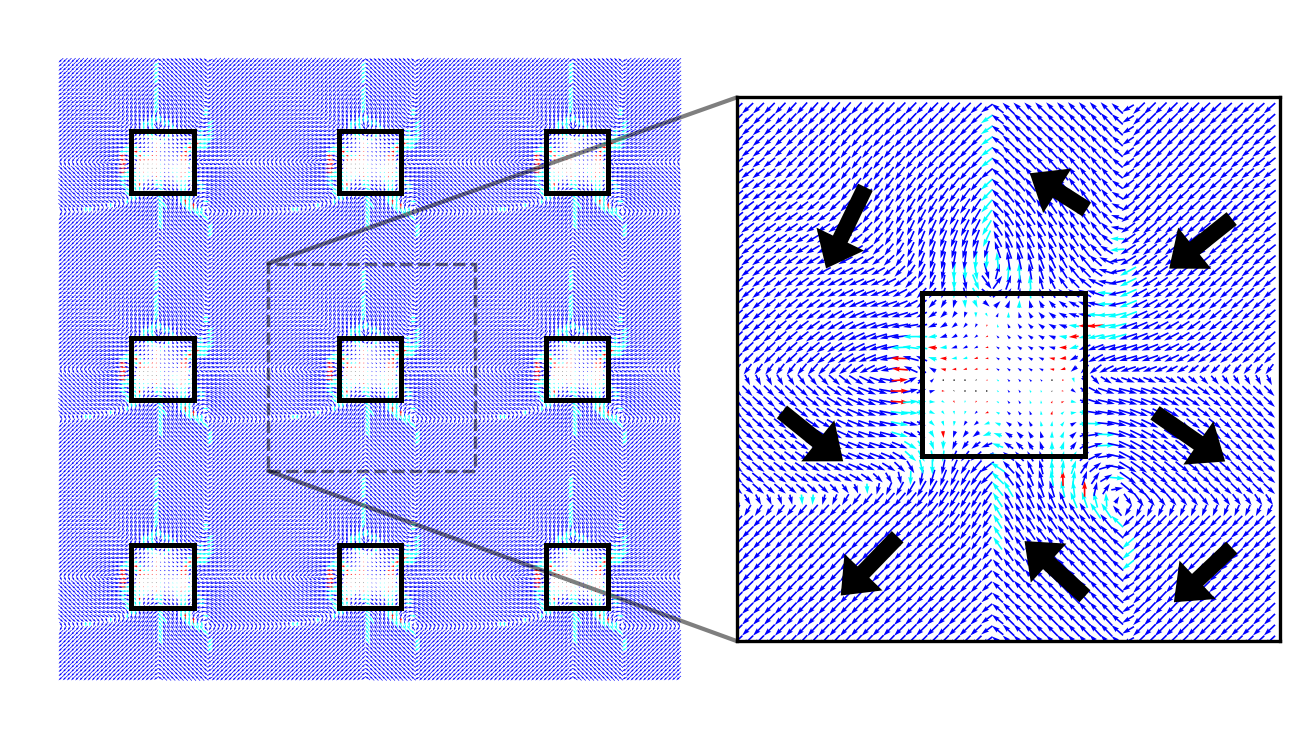}}
        \put(10,115){\scriptsize(c)}
    \end{Overpic}
    }
    \caption{(a) Simulation cell: A cubic inclusions of \ch{Sr} of length $L$ (blue frame) is added in the center of a cubic \ch{Ba} matrix with $N$~u.c.\ length. (b) Illustration of the coarse-graining: All degrees of freedom in each u.c.\ (at position $\textbf{R}$) are mapped on local dipole moments $u(\textbf{R})$, local strain $w(\textbf{R})$ and the local pressure $s(\textbf{R})$ which depends on the local Sr concentration.
    (c) Periodic boundary conditions and domain structure: Overall a composite with equidistant spacings of inclusions is modelled and the local dipoles are characterized with respect to their phase: R (blue), O (cyan), T (red), and C (black) and black arrows illustrate the domain structure. The example shows one representative domain configuration found in cooling simulations for $L=16$~u.c.\ and $N=50$~u.c. (i.e.\ a volume fraction of \ch{Sr} of 3.28~\%) at 150~K.}
    \label{fig:feram_mapping}  
\end{figure}

We utilize the effective Hamiltonian by Zhong et al \cite{zhongPhaseTransitionsBaTiO31994, zhongFirstprinciplesTheoryFerroelectric1995} for solid-solutions of (Ba,Sr)TiO$_3$ with first-principles derived parameters from \cite{nishimatsuMolecularDynamicsSimulations2016} 
\begin{align}
	H^{\text{eff}}&=
	V^{\text{self}}(\{\bm{u}\})+V^{\text{dpl}}(\{\bm{u}\})+V^{\text{short}}(\{\bm{u}\}) \nonumber\\
	&+V^{\text{elas,homo}}({\eta_1,\eta_2,...,\eta_6})+V^{\text{elas,inho}}(\{\bm{w}\}) \nonumber\\
	&+V^{\text{coup,homo}}(\{\bm{u}\},{\eta_1,\eta_2,...,\eta_6})+V^{\text{coup,inho}}(\{\bm{u}\},\{\bm{w}\}) \nonumber\\	
  &+V^{\text{mod,inho}}(\{\bm{w}\},\{s\})\nonumber\\ 
 &+\frac{M_{\text{dipole}}^{*}}{2}\sum_{\bm{R}}\dot{u}_i^2(\bm{R}) 
 -Z^{*}\sum_{\bm{R}}\bm{E}\cdot \bm{u}(\bm{R})\;,
	\label{eq:effham}
\end{align}
where all degrees of freedom are coarse grained to $\{\bm{u}\}$, $\{\bm{w}\}$ and $\{{s}\}$, the soft mode and acoustic displacement vectors, and a measure for the  Sr concentration in each unit cell  (u.c.), respectively, see Fig.~\ref{fig:feram_mapping}.  
The soft mode of each cell is related to its local polarization ($\bm{p}$) as $\bm{p}=Z^{*}\bm{u(\bm{R})}/\Omega$, with $\Omega$ and $Z^{*}$ being its volume and  effective Born charge,\footnote{Following Ref.~\cite{nishimatsuFirstprinciplesAccurateTotal2010} the effective Born charge of the soft mode is given by the sum of the atomic Born charges in the unit cell weighted with their displacement in the corresponding phonon eigenmode.} and $\eta_{1},..., \eta_{6}$ are the six components of the homogeneous strain in Voigt notation ($\eta_1=e_{xx}$, $\eta_4=e_{yz}$).
The potential energy terms are $V^{\text{self}}(\{\bm{u}\})$, the self-energy of the local soft modes, $V^{\text{dpl}}(\{\bm{u}\})$ and $V^{\text{short}}(\{\bm{u}\})$, their long-range and short-range interactions,  $V^{\text{elas,homo}}(\eta_1,\eta_2,...,\eta_6)$ and $V^{\text{elas,inho}}(\{\bm{w}\})$, the elastic energies from homogeneous and inhomogeneous strain, $V^{\text{coup,homo}}(\{\bm{u}\},\{\bm{w}\})$ and $V^{\text{coup,inho}}(\{\bm{u}\},\{\bm{w}\})$, the coupling between the local soft modes and homogenous and inhomogenous strain, and $V^{\text{modulation,inho}}(\{\bm{w}\},\{s\})$,  the coupling between local strain and Sr-concentration. 
As the strain is internally optimized, only the kinetic energies of the local soft mode, with an effective mass $M^{*}_{\text{dipole}}$, is explicitly included in MD simulations and the last term allows to apply an external electric field ($\bm E$).

Following the findings in Ref.~ \cite{walizerFinitetemperaturePropertiesBa2006} that other changes of the energy expansion with Sr concentration have a minor impact on the material description, we use one set of mean parameters and mimic the local Sr-concentration by local strain and a semi-empirical pressure term to address the coupling between $\eta_i$ and $s$. This model can successfully describe the changes of transition temperatures, lattice parameters and polarization with Sr-concentrations and gives us access to interface strain and depolarization fields \cite{nishimatsuMolecularDynamicsSimulations2016}. The potential local enhancement of the Sr-off-centering at the direct \ch{Ba} interface is however neglected \cite{dimouInitioBasedStudyAtomic2024}. 

We model a periodic array of cubic \ch{Sr} inclusions of size $L\times L\times L$~u.c.\ in a matrix of $N\times N\times N$~u.c. of \ch{Ba} using periodic boundary conditions, see Fig.~\ref{fig:feram_mapping}. 
Unless stated otherwise, we use $N=50$~u.c.
MD simulations at finite temperatures are performed using the Nos\'{e}-Poincar\'{e} thermostat and a timestep of 2~fs. Thereby, the systems are thermalized for 40~ps and thermal averages are recorded over 40~ps.
We initialize random dipoles at 400~K or 100~K and simulate heating and cooling with temperature steps of 5~K by taking the final configuration at one temperature as the next initial state. For $L=$ 6, 10, 16, and 28~u.c., at least 5 independent dipole configurations have been analyzed and only for the medium sized inclusion with $L=$16~u.c.\ a variance of the order parameter up to 5~$\mu$C/cm$^2$ is observed.


In order to analyze domain and phase fraction, we average all local dipoles $\bm{p}$ (including matrix and inclusion) over 40~ps and classify them based on 
their polarization components $i,j,k$ relative to the threshold $|p^t|=3.915\mu$C/cm$^{2}$ ($|u^t|$=0.015~\AA) as
\begin{itemize}
    \item cubic (C):~~~~~~~~~~~\,  $|p_i| <|p^t|\wedge |p_j| <|p^t|\wedge |p_k| <|p^t|$ 
    \item tetragonal (T):~~~~\,\, $|p_i| > |p^t| \wedge |p_j|<|p^t|\wedge |p_k| <|p^t|$
    \item  orthorhombic  (O):\, $|p_i|> |p^t| \wedge |p_j|> |p^t| \wedge |p_k|<|p^t|$ 
    \item  rhombohedral (R): $|p_i| >|p^t|\wedge |p_j| >|p^t|\wedge |p_k| >|p^t|$\;.  
\end{itemize}
The macroscopic total polarization $P_i=\langle p_i\rangle$ with $i: x, y, z$ is then given as the ensemble avarage of these dipoles. 
Note that this macroscopic polarization is not a suitable order parameter in the presence of domains.
Instead we use $P^{\text{abs}}=\langle |P_x|\rangle+\langle |P_y|\rangle+\langle |P_z|\rangle$, the ensemble average of the norm of the polarization components,  to identify the phase transitions.
It should be noted that this variable overestimates the macroscopic polarization in each domain.

To identify size and average polarization of domains (3D regions of connected nearest neighbor cells) in the time-averaged dipole configurations, we use the cluster analysis package $cc3d$ \cite{silversmithCc3dConnectedComponents2021} and filter domains with more than 1000 connected unit cells.
For large inclusions, the polarization drops towards their center and we determine the penetration depth as the distance to the interface at which the polarization has dropped by $1/e$. 
All Python scripts and Jupyter notebooks used for the analysis are available on GitLab \cite{tengICAMSSFCSTO_inclusionGitLab2024}.

\section{Results}
\label{sec:Results}
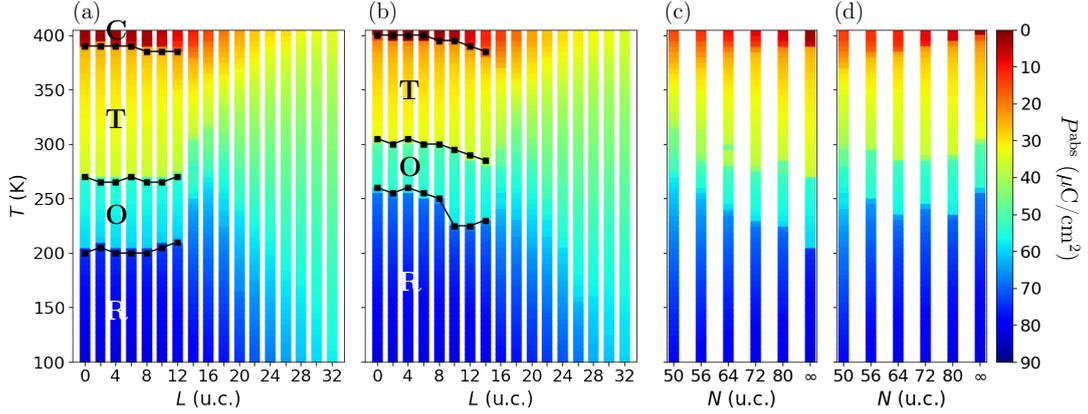
\begin{figure}[t]
    \centering
    \resizebox{\linewidth}{!}{
    \begin{Overpic}[abs]{\begin{tabular}{p{.8\textwidth}}\vspace{.3\textheight}\\\end{tabular}}
        \put(0,0){\includegraphics[height=.32\textheight,clip,trim=0.1cm 0 3.2cm 0]{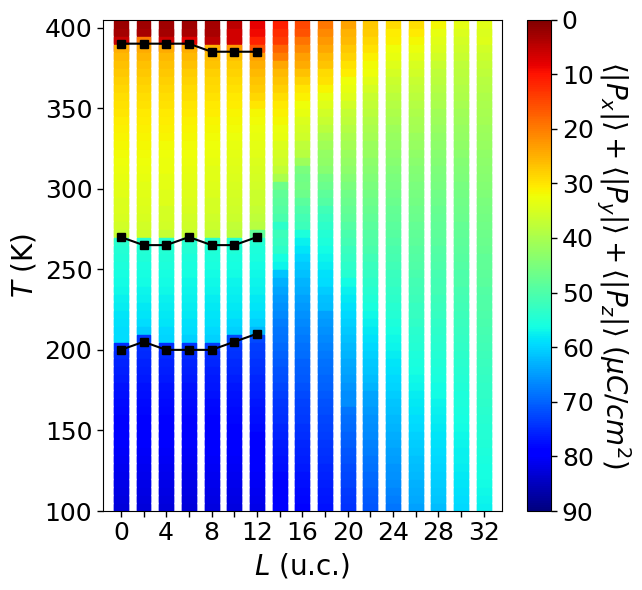}}
        \put(52,196){\Large\textbf{C}}
        \put(52,150){\Large\textbf{T}}
        \put(52,100){\Large\textbf{O}}
        \put(52,50){\color{white}\Large\textbf{R}}
        \put(35,207){\large(a)}
        \put(185,0){\includegraphics[height=.32\textheight,clip,trim=2.5cm 0 3.2cm 0]{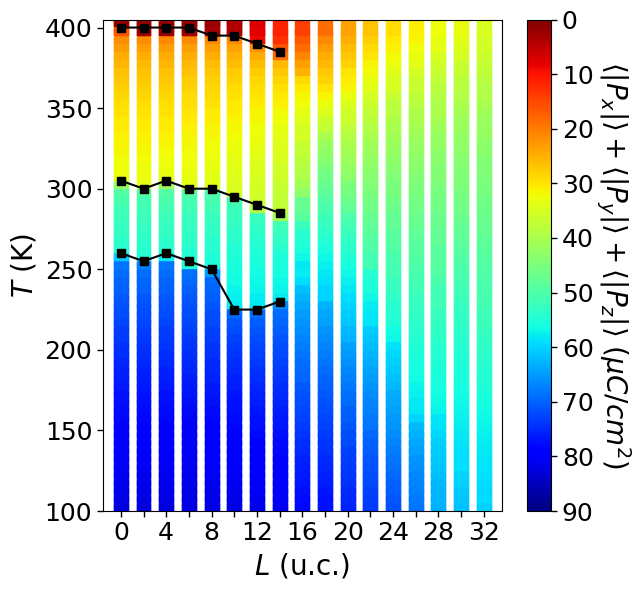}}
        \put(204,165){\Large\textbf{T}}
        \put(204,125){\Large\textbf{O}}
        \put(204,65){\color{white}\Large\textbf{R}}
        \put(188,207){\large(b)}
    \end{Overpic}
    \begin{Overpic}[abs]{\begin{tabular}{p{.5\textwidth}}\vspace{.3\textheight}\\\end{tabular}}
        \put(0,0){\includegraphics[height=.32\textheight,clip,trim=2.5cm 0 3.1cm 0]{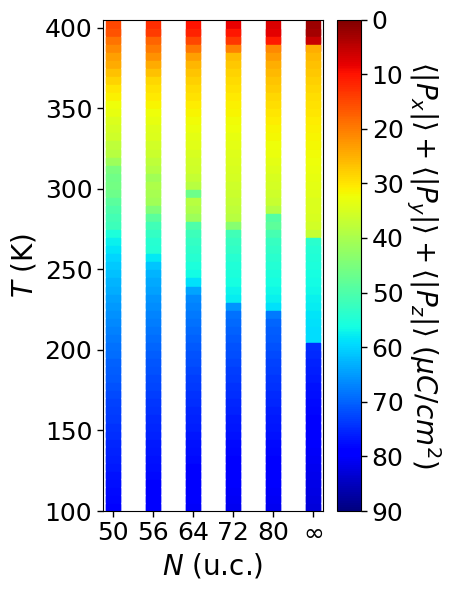}}
        \put(0,207){\large(c)}
        \put(88,0){\includegraphics[height=.32\textheight,clip,trim=2.5cm 0 1.2cm 0]{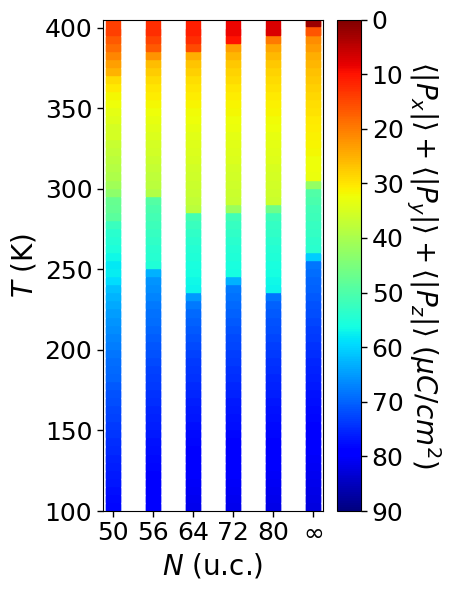}}
        \put(203,115){\rotatebox[origin=c]{270}{\large$P^{\text{abs}}$ ($\mu$C/cm$^{2}$)}}
        \put(88,207){\large(d)}
    \end{Overpic}
    }
    \caption{Phase diagrams: Change of polarization $P^{\text{abs}}=\langle |P_x|\rangle+\langle |P_y|\rangle+\langle |P_z|\rangle$ with temperature  in (a)/(c) cooling and (b)/(d) heating simulations. (a)--(b) Variation of inclusion size, $L$, for a system size of $N=50$~u.c. and (c)--(d)  Variation of system size, $N$, for $L=16$~u.c.
    For $L\leq 8$~u.c., the system is either in single domain cubic (C), tetragonal (T), orthorhombic (O) or rhombohedral (R) phase and black lines give the transition temperatures determined from $d\langle P_i\rangle/dT$. In (c)--(d) $N=\infty$ refers to the pristine material as reference.}
    \label{fig:psum_hc}
\end{figure}

For pristine \ch{Ba} our model predicts three first-order phase transitions from the paraelectric cubic (C) to single-domain tetragonal (T), orthorhombic (O), and rhombohedral (R) ferroelectric phases which are clearly visible as abrupt jumps of $P^{\text{abs}}$ with $\Delta P^{\text{abs}}\approx 17$~$\mu C/cm^2$ at 390~K (400~K), 270~K (305~K), and 200~K (260~K) under cooling (heating) in good agreement with experiments \cite{acostaBaTiO3basedPiezoelectricsFundamentals2017}. 
If Sr inclusions are added for a fixed system size, one may distinguish three different regimes: Small, medium and large inclusions. On the one hand, for small inclusions with $L\leq 8$~u.c., i.e.\ up to 0.41~\% of \ch{Sr}, changes of transition temperatures and thermal hysteresis are not exceeding $\pm5$~K,  see Fig.~\ref{fig:psum_hc}~(a)--(b).
Furthermore, the change of polarization along all directions is within $\pm$0.5~$\mu$C/cm$^2$ and the differences to a homogeneous solid solution of \ch{(Ba,Sr)} with the same Sr concentration are small.
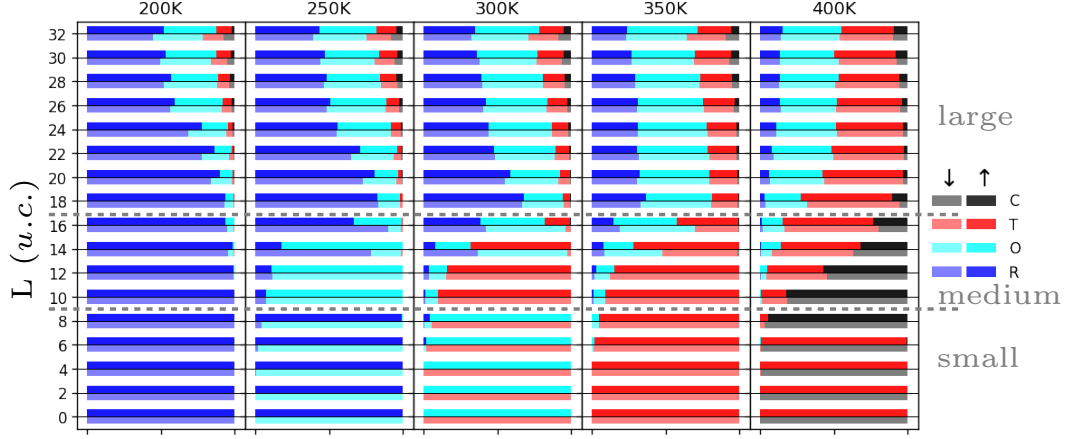
\begin{figure}[t]
    \centering
    \resizebox{1\linewidth}{!}{
    \begin{Overpic}[abs]{\begin{tabular}{p{.5\textwidth}}\vspace{.12\textheight}\\\end{tabular}}
        \put(6,0){\includegraphics[width=.5\linewidth,clip,trim=0 1.7cm 0 0]{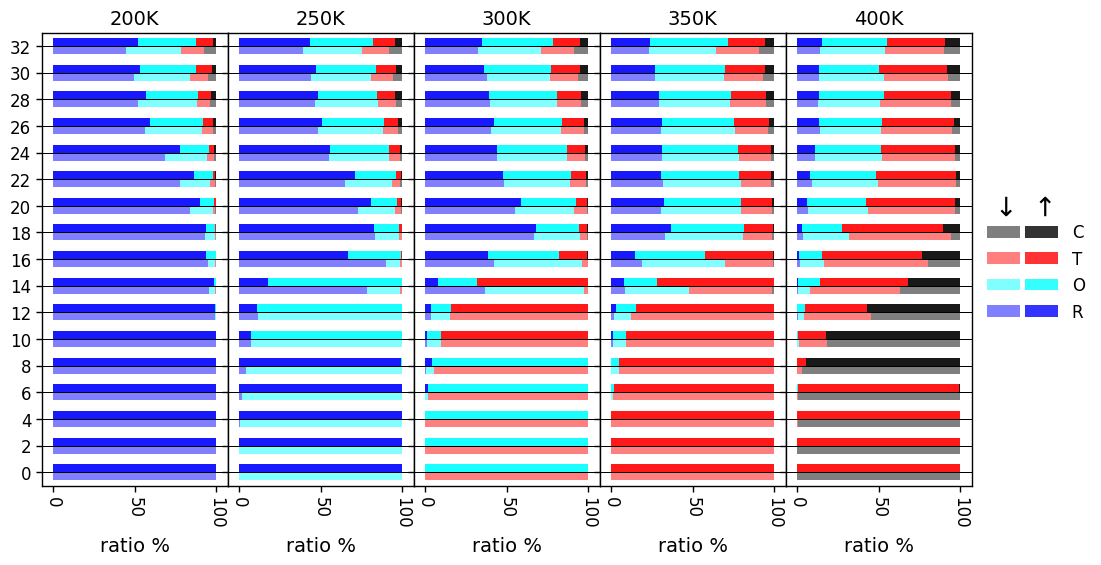}}
        \put(0,40){\rotatebox[origin=c]{90}{\tiny\text{L~($u.c.$)}}}
        \multiput(8,27)(2.5,0){75}{\linethickness{0.25mm}\color{gray}\line(1,0){1.25}}
        \multiput(8,46.5)(2.5,0){75}{\linethickness{0.25mm}\color{gray}\line(1,0){1.25}}
        \put(190,15){\color{gray}\tiny\text{small}}
        \put(190,28){\color{gray}\tiny\text{medium}}
        \put(190,65){\color{gray}\tiny\text{large}}
    \end{Overpic}
    }
    \caption{Phase fraction: Dependency of the fraction of local dipoles classified as C, T, O or R phase on temperature (left to right) and size of \ch{Sr} inclusion ($L$, bottom up) for $N=50$~u.c. For each data set, the upper and lower bars represent heating and cooling simulations. Dashed lines separate: small, medium size and large inclusions as discussed in the text. Note that for $L=8$~u.c., $T_C$ for T--C transition changes from 400 to 395~K, only.}
    \label{fig:ph_comp}
\end{figure}
On the other hand, the homogeneous solid solution and the system with inclusions differ considerably for large inclusions ($L \geq 18$~u.c.). In the former case our model predicts three first-order phase transitions at about 176~K, 224~K, 306~K  for 26.21~\% \ch{Sr} and the polarization is reduced compare to \ch{Ba} in all phases \cite{nishimatsuMolecularDynamicsSimulations2016}. For a corresponding inclusion size of  32~u.c., no signature of any phase transition is visible in $P^{\text{abs}}$ neither for cooling nor for heating. Instead, the changes with temperature are continuous and polarization with a moderate magnitude is stabilized in the whole temperature range. These findings are in agreement with the diffuse phase transitions reported for \ch{Ba}--\ch{(Ba,Sr)}  composites \cite{takeuchiPreparationBaTiO3SrTiO32003}.

For medium-sized inclusions, $8$~u.c.$<L\leq 16$~u.c., the composite shows a transition between both regimes.
For $ L \leq 12$~u.c., the material is still predominantly either in R, O or T phase, separated by first-order phase transitions with thermal hysteresis. 
With increasing $L$, this hysteresis is reduced, and then the transitions become gradually more diffuse. 
These changes of hysteresis are related to an abrupt (R to O) or continuous (otherwise) increase of the transition temperatures under heating. The temperatures where $P^{\text{abs}}$ under cooling reaches the values of T, O and R phases of the pristine material, is minimal or maximal for $L=16$~u.c., for the upper or both lower transitions, respectively.  For this size of the inclusion, phase coexistence is already present even at low temperatures. 
Figure~\ref{fig:psum_hc}~(c)--(d) show the changes of the phase diagram with system size, $N$, for $L=16$~u.c. Under cooling, the changes of all three transition temperatures induces by the inclusion decrease with $N$ and the transitions become less diffuse. Thus, if inclusions of this size have a distance of 6.8~nm, or in other words correspond to 3~\% \ch{Sr} for $N=50$~u.c., they induce large changes of phase diagram and polarization. If their distance increases to 12.8~nm, i.e.\ if the \ch{Sr} the concentration decreases to  0.8~\% for $L=80$~u.c., their impact is small. Under heating, the trends for the lower transitions are less obvious due to different nucleated domain structures.

In the following we restrict the discussion to $N=50$~u.c. As shown in Fig.~\ref{fig:psum_hc}~(a)--(b) $P^{\text{abs}}$ is only reduced with $L$ at low temperatures,  while it is enhanced at and above room temperature.
Note that $P^{\text{abs}}$ is a measure for the magnitude of the local polarization, while the magnitude of the macroscopic polarization, $|\bm P|$ is reduced by large inclusions due to the formation of domain walls as discussed below.    The temperature dependency of $P^{\text{abs}}$ in the presence of large inclusions is related to phase coexistence, see Fig.~\ref{fig:ph_comp}. 
For $L=$32~u.c., all three ferroelectric phases coexists at all temperatures.
Even at 200~K and 400~K, i.e.\ well in the R phase and at the T--C transition of the pristine material, only 52\% or 35\% of the dipoles are in the R or T phase, respectively.
With temperature or $L$, the phase fraction and thus $P^{\text{abs}}$ change gradually. The lower the temperature, the larger is the fraction of the R phase and vice versa for T phase. 

\begin{figure}[t]
    \centering
    \includegraphics[width=.45\textwidth]{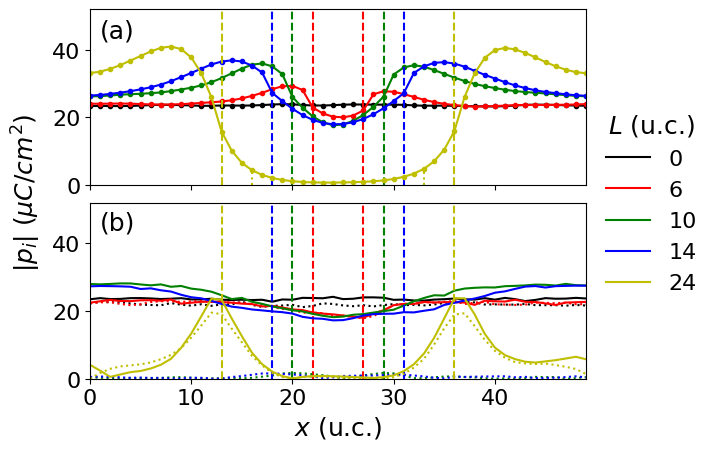}
    \caption{Polarization profiles: Line-scans of local polarization components $|p_i|$ across the center of the inclusions along $x$ at 250~K (heating) for (a) $i:x$ and (b) $i:z$ (solid line), $i:y$ (dotted line)  for different inclusion sizes $L$ (colors). Dashed vertical lines indicate the interfaces between matrix and inclusions.} 
    \label{fig:line_scan}
\end{figure}

How can one understand these changes of phase mixtures and character of the phase transitions? As shown for the example of 250~K in Fig.~\ref{fig:line_scan}, small inclusions are fully polarized by the surrounding matrix in agreement to predictions for superlattices \cite{neatonTheoryPolarizationEnhancement2003}. In turn phase stability and polarization in the matrix are not modified, only the polarization normal to the inclusion increases locally by the compressive interface strain as discussed below.
The polarization in the inclusion however drops with the interface distance, for example by 20~\% at the center of an inclusion with $L=6$~u.c., see red lines in Fig.~\ref{fig:line_scan}. The core of large inclusions is  paraelectric with a penetration depth of about 3~u.c., see yellow lines in Fig.~\ref{fig:line_scan}. 
The resulting polarization gradients induce depolarization fields $E^{\text{dep}}\sim\nabla \cdot P_i$ along the normals of the inclusion \cite{dawberPhysicsThinfilmFerroelectric2005} which destabilize homogeneous polarization along $i$.
\footnote{Note that the inclusions of intermediate size  shown in green and blue in Fig.~\ref{fig:line_scan},   stabilize the orthorhombic phase at 250~K. Thus the two polarization components ($p_z=p_x$) in the center of the \ch{Ba} matrix increase realative to the pristine rhombohedral phase.}
For $L=24$, flux-closure domain structures are induced, in agreement to previous reports on \ch{Pb}/\ch{Sr} superlattices \cite{yadavObservationPolarVortices2016}, whose mean in-plane polarization at the interface is as large as the polarization in the rhombohedral phase of \ch{Ba}.
In addition, the depolarization field induces local polarization rotation at the edges of the inclusions in the regime of small $L$. This results in a small phase fraction of O or R dipoles shown in Fig.~\ref{fig:ph_comp} already at 350~K and 250~K, respectively. 

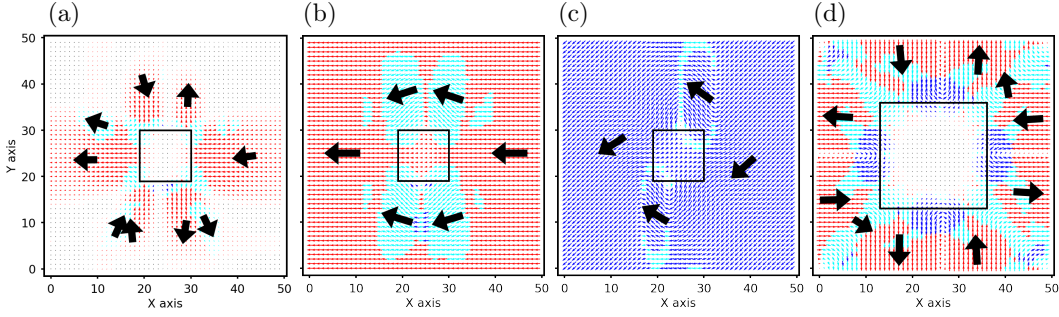
\begin{figure}[t]
    \centering
    \resizebox{1.\linewidth}{!}{
    \begin{Overpic}[abs]{\begin{tabular}{p{1.\textwidth}}\vspace{.165\textheight}\\\end{tabular}}
        \put(1,0){\includegraphics[height=.18\textheight,clip,trim=0cm 0cm 0cm 0cm]{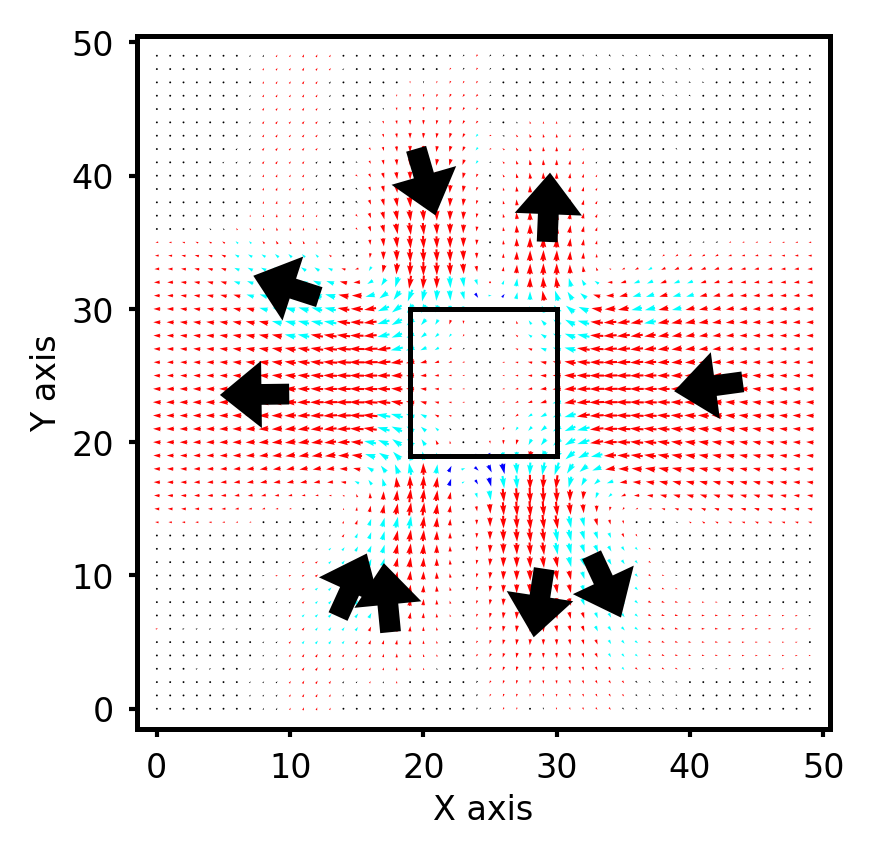}}
        \put(21,118){\small(a)}
        \put(120,0){\includegraphics[height=.18\textheight,clip,trim=1.1cm 0cm 0cm 0cm]{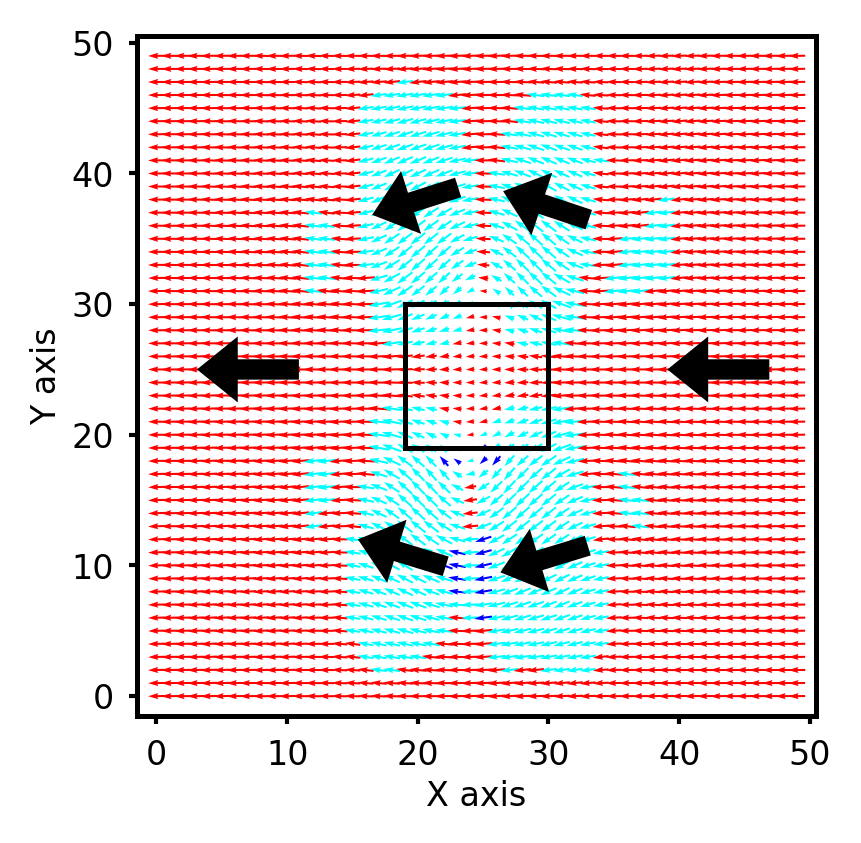}}
        \put(121,118){\small(b)}
        \put(220,0){\includegraphics[height=.18\textheight,clip,trim=1.1cm 0cm 0cm 0cm]{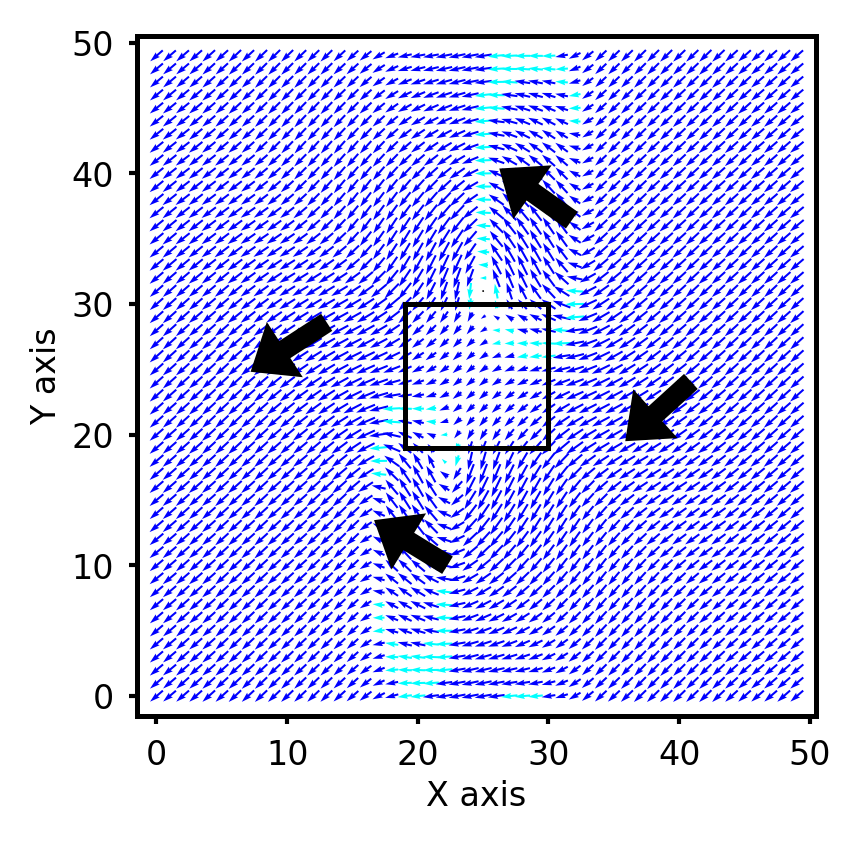}}
        \put(221,118){\small(c)}
        \put(320,0){\includegraphics[height=.18\textheight,clip,trim=1.1cm 0cm 0cm 0cm]{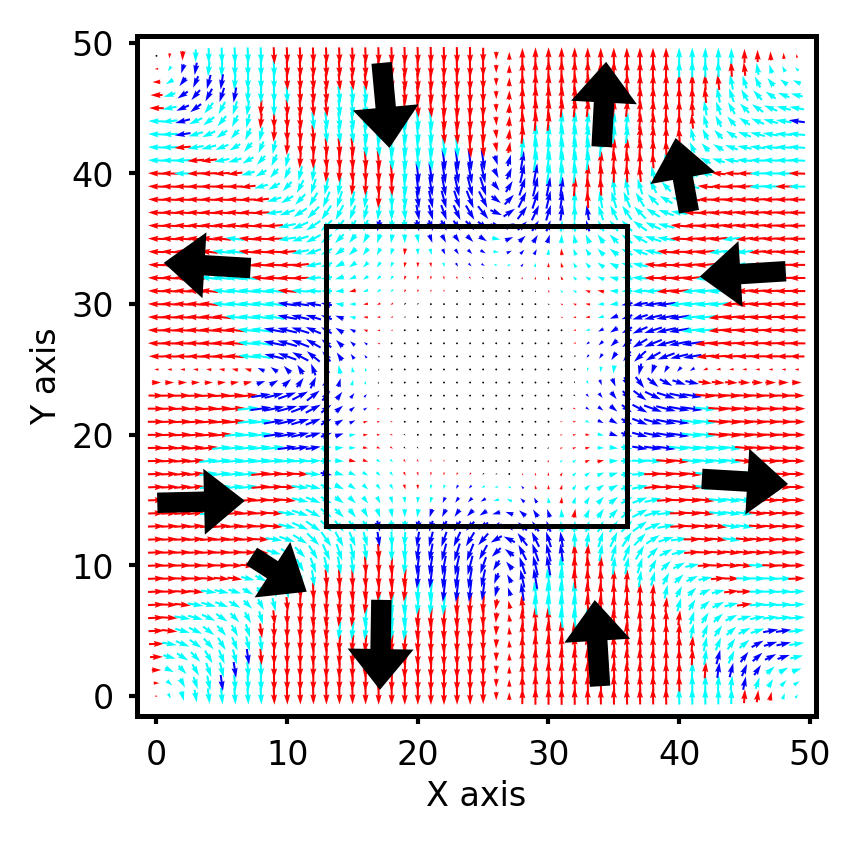}}
        \put(321,118){\small(d)}
    \end{Overpic}
    }
    \caption{Dipole patterns: Cross-sections of time-averaged local dipoles in the center \ch{Sr} inclusion  ($z=25$) for (a)--(c) the medium inclusion size $L=12$~u.c. and (d) the large inclusion $L=24$~u.c. at (a) 400 K, (b) 300 (c) 200~K, and (d) 350~K during heating simulations.  Colors indicate dipole classes (R: blue, O: cyan, T: red) and black arrows mark the direction of polarization in chosen domains. Note that $\pm p_y$ domains also occur at the not shown $x$-interface of the inclusion in (a) and (d) and that qualitatively the same dipole patterns at the interfaces as in (a)--(c) are found for $L=16$~u.c.\ and $N=80$~u.c.}
    \label{fig:cross}
\end{figure}

For $L\geq12$~u.c., a multidomain phase with T 180$^{\circ}$ walls along the surface normal is induced at 400~K, i.e.\ in he C--T phase coexistent temperature range of the pristine material, see Fig.~\ref{fig:cross}~(a).  
At this size, also the transition temperature between the T-like and C-like phases drops and the change of $P^{\text{abs}}$ with temperature becomes diffuse, see Fig.~\ref{fig:psum_hc}. Furthermore, domains are stabilized in the upper ferroelectric phases by inclusions with $L\geq 14$ or 16~u.c.\ during heating or cooling simulations, respectively. Thereby the number of tetragonal domains at 400~K is about 6,  and for large inclusions all possible variants of the T phase ($\pm x$,  $\pm y$ and $\pm z$) occur, see Fig.~\ref{fig:cross}~(d). Note that tetragonal $\pm p_y$ domains form analogously on the not shown interface of the inclusion.

The critical size for the formation of domains in the R phase is $L=10$~u.c. With $L$ first, bubble domains separated by 71$^{\circ}$  walls along the surface normal $i$ and a local reversal of $p_i$  are induced during heating, see Fig.~\ref{fig:cross}~(c). 
These bubbles promote the transition to the O phase and thus abruptly reduce the thermal hysteresis of the O--R transition.
For $L=16$~u.c.\ also stripe domains and different superpositions of 71$^{\circ}$ and  109$^{\circ}$ walls become metastable. One representative example is shown in  Fig.~\ref{fig:feram_mapping}~(c). During cooling the fraction of 109$^{\circ}$ walls is typically smaller compared to heating.
In the large $L$ regime, a heterophase is induced, which is composed of dense nano-sized domains, see Fig.~\ref{fig:cross}~(d), and we detect more than 20 different domains larger than 1000~u.c.\ for $N=50$~u.c.\ and $L\geq 18$~u.c. In this case the inclusions span already half of the system dilation and are thus large obstacles for the formation of the equilibrium domain structure.

In addition to depolarization, the \ch{Sr} inclusions induce compressive interface strain due to the lattice mismatch of 2.1~\%. This strain and the strained interface area increase with $L$. Analogous to the increase of $p_i$ along the direction of the interface normal $i$ in compressively strained films \cite{grunebohmInitioPhaseDiagram2015}, $p_i$ at the interface thus increases, see Fig.~\ref{fig:line_scan}, e.g.\ by 50~\%  for $L=14$~u.c. 
For small and medium sized inclusions, $p_i$ drops to the pristine value well in the matrix, while 
$p_i$ is enhanced in the whole \ch{Ba} matrix for larger volume fractions of \ch{Sr}. This increase of $p_i$ at the interfaces stabilizes tetragonal domains in a broad temperature range also above the transition temperature of the pristine material for $L>14$~u.c. At lower temperatures, 
the O phase starts to nucleate at the edges of the inclusion, see Fig.~\ref{fig:cross}~(b) and is stabilized there in a broad temperature range. Thereby, 
$[110]$-type domains at $[001]$-type edges are favorable as they do not induce large depolarization fields and are compatible with the strain imposed by the adjacent [100] and [010]  faces. 
Analogously, the R phase starts to nucleate at the edges of the inclusion for small and intermediate $L$. In contrast to phase fraction and macroscopic polarization discussed above, the dipole patterns induced at the interfaces of the inclusions are not sensitive to the distances of neighboring inclusions. Even for  $N=80$~u.c.\ i.e.\ distances of 12.8~nm between inclusions, qualitatively the same dipole patterns as in Fig.~\ref{fig:cross}~(a)--(c) are induced. 
In case of large inclusions, the strain stabilizes tetragonal domains perpendicular to all interfaces, i.e. $\pm x$, $\pm y$ and $\pm z$ at x,y and z interfaces, and the enhanced polarization along $i$ at the faces of the inclusion is furthermore screened by flux-closure dipole pattern which are predominantly of R-type directly at the interface, see Fig.~\ref{fig:cross}~(d).

In summary, the local polarization at the interfaces to the inclusions increases with its size and a fraction of the lower temperature phase(s), e.g. \ T in C phase, or O and R in T phase, is already stabilized at higher temperatures resulting in a heterogeneous state with coexisting phases and domains for large inclusions. With temperature the phase fraction changes continuously and thus the first-order character of the whole matrix is lost for few \% of Sr in the system.\\

\begin{figure}[t]
    \centering
    \resizebox{.9\linewidth}{!}{
    \begin{Overpic}[abs]{\begin{tabular}{p{.9\textwidth}}\vspace{.135\textheight}\\\end{tabular}}
        \put(0,0){\includegraphics[height=.15\textheight,clip,trim=0 0 3.2cm 0]{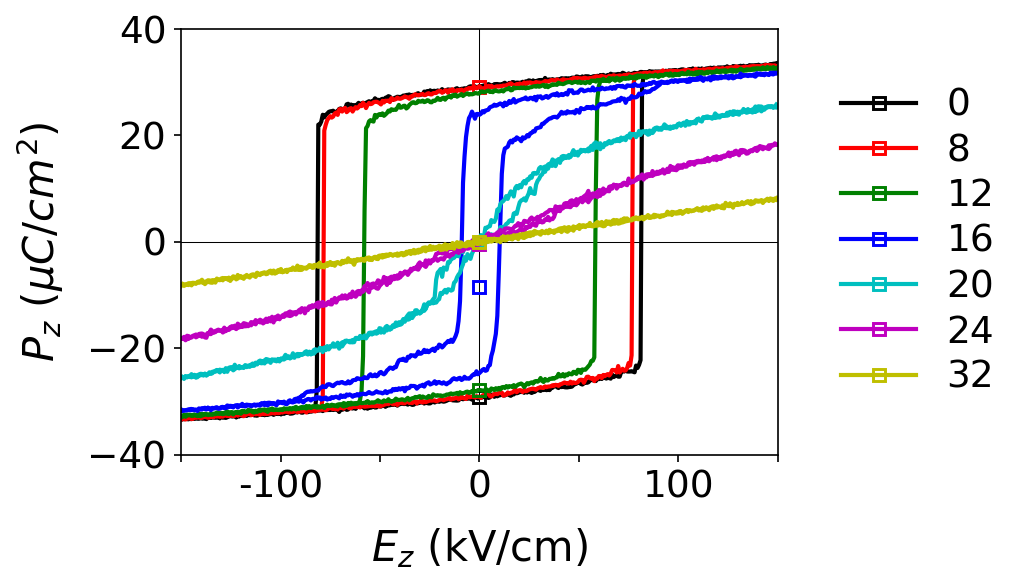}}
        \put(32,85){\scriptsize(a)}
        \put(134,0){\includegraphics[height=.15\textheight,clip,trim=2.8cm 0 3.2cm 0]{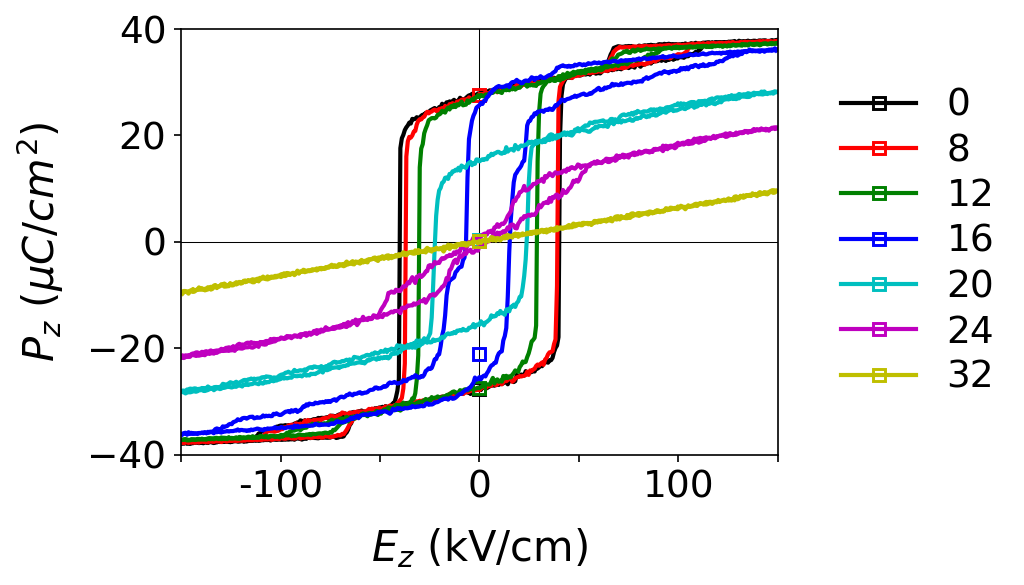}}
        \put(139,85){\scriptsize(b)}
        \put(241,0){\includegraphics[height=.15\textheight,clip,trim=2.8cm 0 0cm 0]{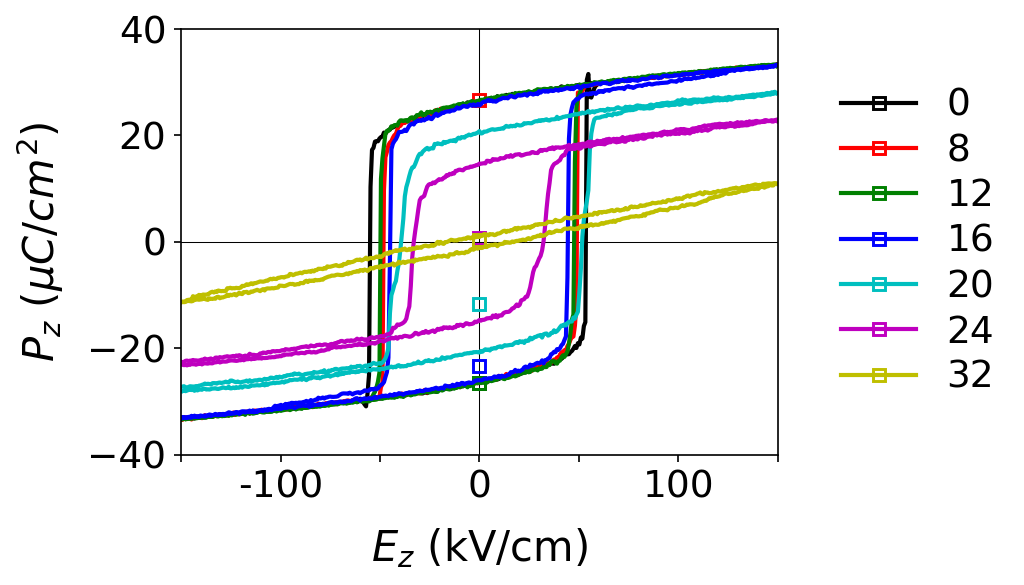}}
        \put(246,85){\scriptsize(c)}
    \end{Overpic}
    }
    \caption{Field hysteresis: Impact of inclusions on the $P_z$ field hysteresis at (a) 350~K, (b) 250~K and (c) 150~K  for $N=50$~u.c.
    The field is applied along $z$ and only the polarization component parallel to the field is given. 
    Colors indicate the pristine material (black), $L=$8~u.c.\ (red), 12~u.c.\ (green), 16~u.c.\ (blue), 20~u.c.\ (cyan), 24~u.c.\ (purple) and 32~u.c (yellow). Dots mark the initial state after cooling and one hysteresis loop after prepoling to $-150$~kV/cm is shown.}
    \label{fig:pehys_tor_snapshots}
\end{figure}
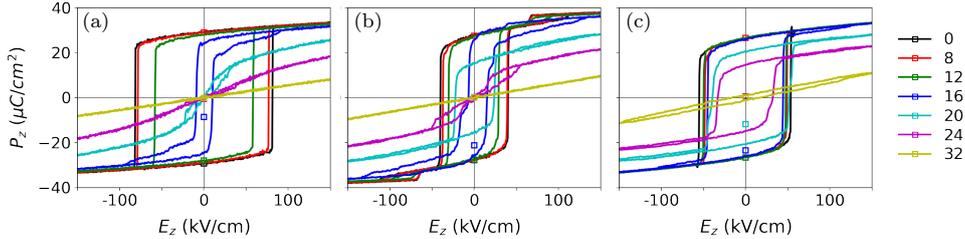
These changes of phase and domain composition with $L$ strongly impact the field-response of the material, see Fig.~\ref{fig:pehys_tor_snapshots}.  As reference, black lines show the responses of the pristine material in (a) tetragonal (b) orthorhombic and (c) rhombohedral phase under a field along [001]. Our model predicts coercive fields of about 82~kV/cm, 41~kV/cm and 54~kV/cm, respectively. Note that at $250$~K furthermore the O to T transition is induced at  $E=\pm$62~kV/cm.
As discussed above, the impact of small inclusions  (green) on the macroscopic polarization is small. Nevertheless, the coercive field is already reduced in all phases, e.g.\  by 10~\% for the O phase by such small inclusions. One may speculate that the polarization rotation at the edges of the inclusion promotes the switching and as the same rotation is induced in systems with larger distances between inclusions, a reduced coercive field is likely also there. 

With increasing size of the inclusion, the coercive field is systematically reduced at all temperatures, e.g.\ by 30~\% for L=$12$~u.c. and by 90~\% for $L=16$~u.c. at 350~K.
Due to the discussed metastable domain configurations for medium sized inclusion, particularly for $L=16$~u.c. in the R phase, the initial state after cooling (dots) may differ from the remanent states during field cycling. However for all configurations, we observe a reversible change in domain and phase fraction during field cycling.
It is important to remember that  $L=16$~u.c.\ corresponds to a volume fraction of 3\% of \ch{Sr}. This  small Sr concentration already has a major impact on the field hystereses, while the overall remanent and saturation polarization of the matrix are similar to the pristine material. 

The multidomain phases without macroscopic polarization found for large and dense inclusions, e.g.\ for $N=50$ and $L=20$~u.c.\ at 350~K and for $L=24$~u.c.\ at 250~K, furthermore allow for thin double loop hysteresis,
However, in these cases $E_z=\pm150$~kV/cm is not sufficient to fully overcome the depolarization field and to switch all dipoles above or below the inclusion (along $z$ direction). Thus the switchable polarization is reduced.
Finally, \ch{Ba} with \ch{Sr} inclusions of $L=32$~u.c.\ and $N=50$~u.c., i.e.\  26.2\% of Sr show a linear dielectric response for all temperatures.


\section{Conclusion}
\label{sec:Conclusion}
We have used coarse-grained MD simulations based on an \textit{ab\ initio} derived effective Hamiltonian to study \ch{Ba}--\ch{Sr} composites with cubic SrTiO$_3$ inclusions in the ferroelectric matrix. For this morphology which has been so far underrepresented in literature, we have screened the effect of the inclusion size up to length of 12.8~nm, i.e.\ $L=32$~u.c., and \ch{Sr} concentrations of 0.1--25\%. 

The calculated  polarization profiles reveal that  phase and domain stability as well as the field response of the composite can be grouped into three different size regimes: 
First, small \ch{Sr} inclusions with lengths below 3.2~nm, i.e. $L\leq8$~u.c., are fully polarized by the surrounding matrix and have a minor effect on phase stability and phase transitions. However, these small inclusions may already reduce the coercive field of the material by local polarization rotation at the edges of the inclusions. 

Second, medium-sized inclusions with lengths ranging between 4 and 6.4~nm, i.e.\ $L=10-16$~u.c., act as nucleation centers for phase transition particularly under heating and thus reduce the thermal hysteresis of the transition.  Furthermore, the coercive field during field-induced switch can be reduced by up to $90$\%. Note that these large changes can be induced by an overall \ch{Sr} concentration of only about 5\% and thus the remanent polarization of the pristine material is conserved. 

Third, dense periodic inclusions with lengths of at least 6.4~nm ($L\geq16$~u.c.), can induce complex multidomain states with phase coexistence in a broad temperature range. The local polarization in these complex heterogeneous states may exceed that of pristine BaTiO$_3$ due to the compressive strain induced by the interfaces. The macroscopic polarization however disappears with increasing size of the inclusions. This allows for pinched double loop hystereses.
Diffuse phase transitions and a continuous change of the phase fraction with temperature rather than clear phase transitions are promising for constant functional properties in a broad temperature range. 

Our results show the high potential of paraelectric inclusions in a ferroelectric matrix to tune phase stability and domain structure. In contrast to superlattices, their morphology impacts the polarization  along all three directions the same way allowing for heterogeneous phases. As the phase nucleation and stability is related to the faces and edges of the inclusions even small changes of the geometry like rounded corners or elongation may allow for large modifications.


\section*{Acknowledgement(s)}

This work was supported by the German research foundation (DFG) GR 4792/3. 
We thank Prof.\ Dr.\ Markus Stricker for fruitful discussions and constructive feedback.

\section*{Disclosure statement}

The authors declare no conflict of interest.

\section*{Declaration of generative AI in scientific writing}

During the preparation of this work, the authors used DeepL and Grammarly for wording suggestions and checking of language. After using these tools, we reviewed and edited the content as needed and take full responsibility for the content of the publication.


\bibliographystyle{tfq}
\bibliography{references.bib}


\end{document}